# Accuracy Improvement Technique of DNN for Accelerating CFD Simulator

Yukito Tsunoda [1]
*Fujitsu Limited, Kawasaki, Kanagawa, 211-8588, Japan*

*University of Tokyo, Bunkyo-ku, Tokyo, 113-8656, Japan*

Toshihiko Mori [2], Hisanao Akima [3], Satoshi Inano [4], Tsuguchika Tabaru [5]
*Fujitsu Limited, Kawasaki, Kanagawa, 211-8588, Japan*

and

Akira Oyama [6]
*Japan Aerospace Exploration Agency, Sagamihara 229-8510, Japan*

For expensive real-world aerodynamic design optimizations, a new approach for acceleration of flow simulations using DNN is proposed. In this approach, plausible initial flow condition is predicted by DNN to significantly reduce number of time steps required for convergence to the steady state. DNN is trained with flow simulations prior to aerodynamic design optimization where coordinates and distance from the wall surface of all grid points for flow simulation are the input and physical properties such as static pressure and flow velocity are the output. In this study, the proposed approach was evaluated for an aerodynamic airfoil shape design optimization problem. The results showed that 3.9x speed up is achieved in average compared with standard flow simulation without initial flow prediction by DNN. Though the proposed approach requires additional flow simulations before aerodynamic design optimization for training DNN, in total, more than 3.64x speed up can be achieved for aerodynamic design optimizations, for example, an aerodynamic design optimization that requires 10,000 flow simulations and 180 additional flow simulations for training of DNN.

## I. Nomenclature

*AoA* = angle of attack
$\rho$ = density nondimensionalized by the density of ambient
*u* = velocity of x direction nondimensionalized by the sound speed of ambient condition
*v* = velocity of y direction nondimensionalized by the sound speed of ambient condition
*e* = total energy nondimensionalized by density and sound speed of the ambient condition
*J* = Jacobian
$C_L$ = lift coefficient

---

[1] Senior Researcher, Fujitsu Limited; tsunoda.yukito@fujitsu.com
[2] Researcher, Fujitsu Limited; mori.toshihiko@fujitsu.com
[3] Manager, Fujitsu Limited; akima.hisanao@fujitsu.com
[4] Researcher, Fujitsu Limited; inano@fujitsu.com.
[5] Project Manager, Fujitsu Limited; tabaru@fujitsu.com.
[6] Associate Professor, Institute of Space and Astronautical Science; oyama@flab.isas.jaxa.jp. Senior Member AIAA.

$C_D$ = drag coefficient
*Loss 1* = loss function used to train DNN consisting of mean square error
*Loss 2* = loss function used to train DNN consisting of customized mean square error
$j$ = index of grid points
$k$ = index of grid points
*jmax* = number of grid points
*kmax* = number of grid points
$A_p$ = representative value on CFD grid inferred by DNN
$A_t$ = ground truth value on CFD grid
$S(j,k)$ = the size of the associated area for each grid point

## II. Introduction

Multi-objective evolutionary algorithms (MOEAs) have been successfully applied in various fields. MOEAs have excellent features, such as the capability to obtain the Pareto optimal solutions. In the field of aerospace engineering, this approach has been employed to optimize the design of space trajectories [1-3], earth observation satellites mission planning [4], design of rocket engines [5-8], flame deflectors [9], and aerodynamics design problems [10-19]. Suitable design candidates were often obtained. In an MOEA, an evaluation of the objective function is required for each design candidate. Therefore, to optimize the aerodynamic design using an MOEA, the computational cost of computational fluid dynamics (CFD) is a serious issue. In case the optimization problems for flame deflectors [9], the computational time required for simulation of each design candidate is approximately 7 h with 130 processors (1040 cores). 2,500 design candidates were evaluated in this research. Therefore, total computational time is still more than 2 weeks although a large-scale supercomputer is used to solve this problem. This enormous computational time is serious issue. So, the computational cost of this simulation to evaluate each design candidate should be suppressed.

One method for reducing the computational cost is to replace the CFD with a surrogate model. There are reports of the use of a surrogate model, such as Kriging [18], a radial basis function [19], or a neural network (NN) [20]. In these reports, surrogate models were utilized to calculate the value of the objective function. Owing to this method, suitable design candidate can be obtained with realistic computational cost. Here, most often, optimization of aerodynamic design tends to be performed with many design variable [21,23]. An optimization performed with more design variables enables to evaluate various design. As a result, a better design is often obtained. [23]. However, for the surrogate model, the number of design variables becomes a critical factor. The number of samples required to estimate a surrogate model of several variables grows exponentially with the number of variables [21, 22]. It is difficult to use surrogate models for design optimization under many design variables.

We suggest accelerating the CFD rather than utilizing a surrogate model. There are techniques to incorporate a deep neural network (DNN) in CFD to accelerate [24-26]. In particular, one method satisfies the rules of physics equations of the original solver [26]. In this method, the procedure of the CFD incorporating the DNN inference is performed as follows: The CFD procedure is performed in some initial steps (warm-up CFD stage). Subsequently, the DNN infers the flow field of the steady solution from the flow field of the warm-up CFD (DNN inference stage). Next, the CFD procedure is performed again from the inferred flow field of the DNN (main CFD stage). The CFD simulations were performed with representative designs in advance. Then, the DNN is trained using these simulation results. With this DNN, some iterative CFD computations are replaced by DNN inference. Consequently, the computational cost is reduced. In addition, the main CFD stage refines the error caused by DNN inference. Therefore, this method succeeded in eliminating the difference between the output of the CFD and the output of the traditional CFD [26].

Here, we would like to address two issues based on the CFD simulation to obtain further improvements.

1: The warm-up CFD stage is performed by starting the CFD simulation from a uniform flow field. However, we anticipate that simulation failures may sometimes occur at this stage. We anticipate that this is caused by a drastic change the flow field occurs at initial stage by starting this uniform flow field. Therefore, we aim to eliminate this warm-up CFD stage.

2: The number of iterations of the main CFD stage depends on the accuracy of the DNN inference. The computational cost largely depends on the number of iterations. Therefore, we attempted to further improve the accuracy of the flow field of the DNN inference.

We propose two techniques to overcome these issues. In the previous research, the flow field calculated by the warm-up CFD is used to input of DNN [26]. Therefore, the warm-up CFD is indispensable with this method. To overcome this issue, the first technique is to directly infer the flow field of the steady solution from the shape of the design candidate. We used the information of the CFD grid instead of the information of flow field. The position on the coordinates of the grid point and the distance from the surface of design were used as the input information of the DNN. This method eliminates the process of the warm-up CFD stage. The second technique is to use a customized mean square error (MSE) as a loss function used for DNN training. In CFD, the evaluation grid points are usually nonuniformly located in the coordinate system. In this case, the grid points located in the sparse region have a large associated region. The influence of the grid points located in the sparse region is proportional to the size of the region. Therefore, the error on the grid point located in the sparse region also has a significant influence proportional to the size of associated region. This influence is not considered in the previous research [26]. Training DNN by considering this influence, the accuracy of the DNN inference will be expected to improve further from the perspective of the entire flow field. Each square error was weighted depending on the size of the associated area because the influence of error is proportional to the size of associated region.

The purpose of this study is to reduce the total computational cost of aerodynamic design optimization problems using steady-state flow simulations. The suppression of the computational cost was realized by suppressing the cost of CFD simulation. To overcome this issue, we proposed new techniques to further improve the CFD simulation accelerated by incorporating DNN inference. Design optimization problem of airfoil shape using PARSEC parameters [17, 18] was used as the sample problem this time. In the design optimization problem, various design candidates are evaluated using CFD simulations. Therefore, various airfoil shapes were created based on the PARSEC parameters [17, 18, 27] and evaluated. The structured grid created depending on a self-adaptive-grid method [28] is used as CFD grid. Then, CFD simulations performing against these airfoil shapes were used to evaluate the computational cost. The total computational cost for design optimization is also estimated using these results.

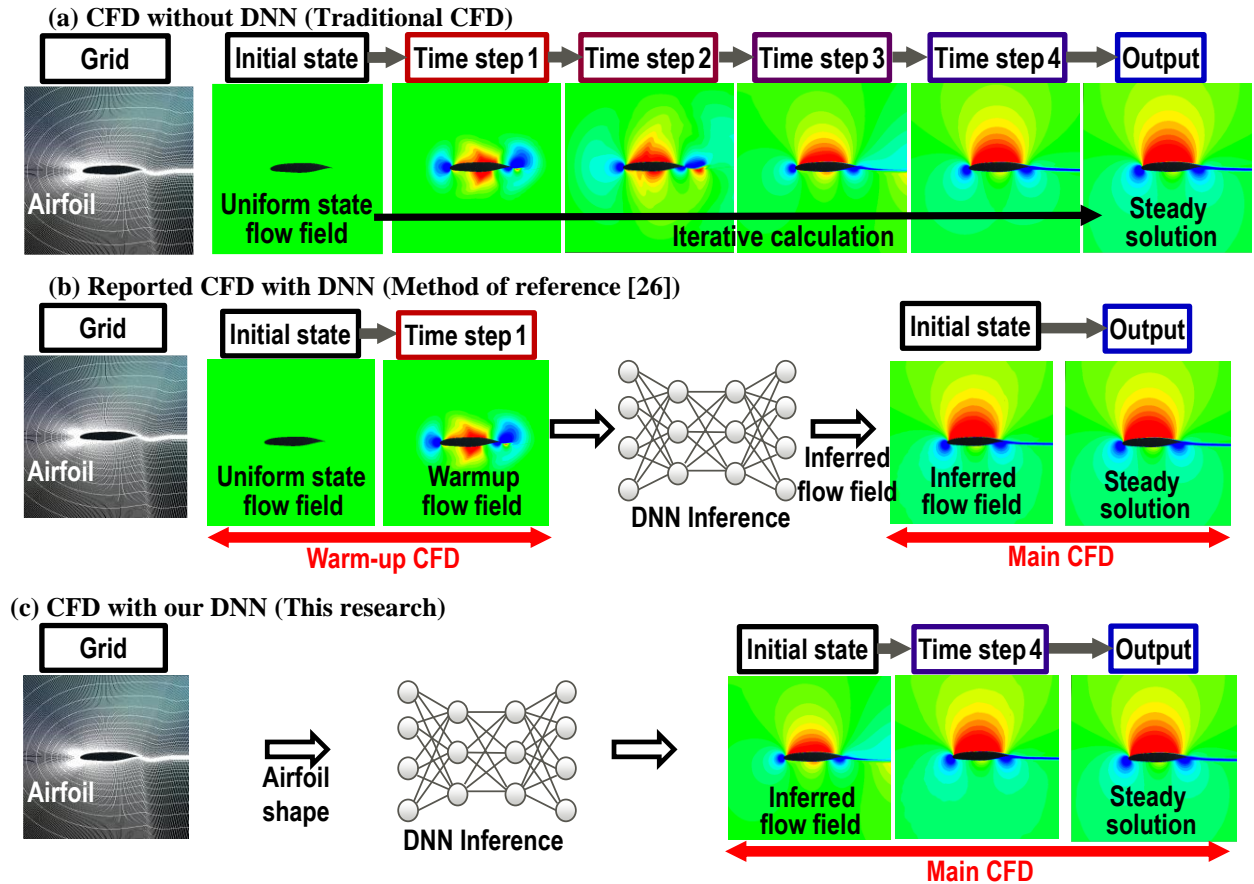

Fig. 1. Process image of the CFD simulator without DNN (a), reported CFD with the DNN (b) and CFD with our DNN. In CFD with our DNN, a steady state flow field is inferred by the DNN, and the CFD is performed from this inferred flow field.

The CFD with the DNN incorporated our first technique was evaluated, first. The evaluation results show that the CFD with the DNN incorporating our first technique achieves 1.7× speedup against the CFD without DNN inference. Besides, we succeed in eliminating warm-up CFD stage used in the CFD in the previous research [26]. The CFD with the DNN incorporated both the first and the second technique was evaluated, next. The evaluation results show that this CFD achieves a 3.9× speedup compared to the traditional CFD. This result shows that further speed up as 2.3× is realized by using the second technique.

DNN training was performed with 180 datasets of CFD grids and the flow fields in this study. Therefore, 180 datasets are created using traditional CFD simulator, first. Then, the DNN is trained using these 180 datasets. It takes 9.5 hour to train the DNN. This computational time is overhead of this method. However, the computational time of CFD against each design candidate is suppressed from 10.5 hour with traditional CFD to 2.7 hour with our CFD. So, this overhead is enough small. For example, in case of design optimization with 1,000 design candidates, the total time is suppressed from 475 hours with traditional CFD to 206.7 hours with our CFD.

### III. Approach to Improve the CFD Incorporating the DNN Inference

We developed a new approach to the accelerated CFD simulation by incorporating DNN inference. The computational cost is suppressed by starting the CFD from a flow field close to the steady solution. This flow field is produced by the DNN inference. The processed image of our CFD, compared with the other reported CFD, is shown in Fig. 1. The shape of the design candidate was utilized in the DNN input. This flow field was created using DNN inference. Some iterative calculations of CFD are replaced by DNN inference. Consequently, the computational cost of the CFD can be suppressed. In addition, the warm-up CFD stage, starting from the uniform flow field, can be eliminated. We anticipate that simulation failure may sometimes occur at this warm-up CFD stage. However, we can avoid this possibility.

#### A. Input Information of DNN

The DNN needs to infer the flow field of the initial state of the CFD from the shape of the design candidate. The structured grid created depending on a self-adaptive-grid method [28] is used this time. In case of using structured grid, the allocation of grid points depends on the shape of the design candidate. Therefore, information about the positions of the grid points also includes information about the shape. For these reasons, the position of each grid point can be applied as input information for the DNN. The procedure for creating the input information is shown in Fig. 2. Both "*x*", "*y*" coordinates and distance from the surface of airfoil "*L*" of each grid point were employed as the input information. The flow state is determined by the relative position of the airfoil. The information about the positions of the "*x*" and "*y*" coordinates is used to determine the relative direction against the airfoil. The distance information "*L*" is used to recognize the relative distance from the surface of the airfoil. This distance information is used such as to recognize whether it is positioned within the boundary layer or not.

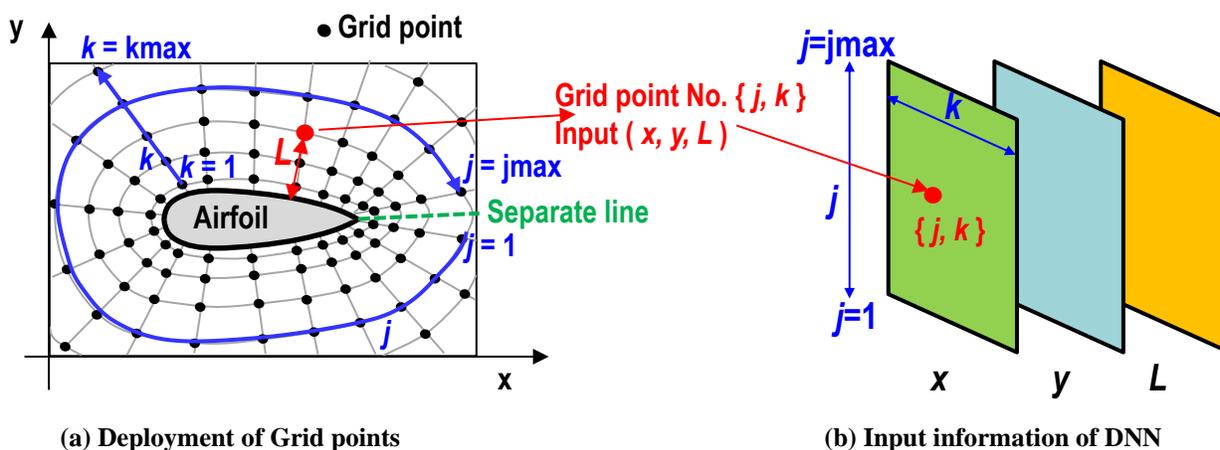

(a) Deployment of Grid points          (b) Input information of DNN

Fig. 2. DNN architecture and input-output information. The grid point of CFD is placed around the evaluation airfoil (a). Grid point {j,k} are treated as the pixel of the image. The variables "x", "y", and "L" of grid point is used as an input of DNN (b).

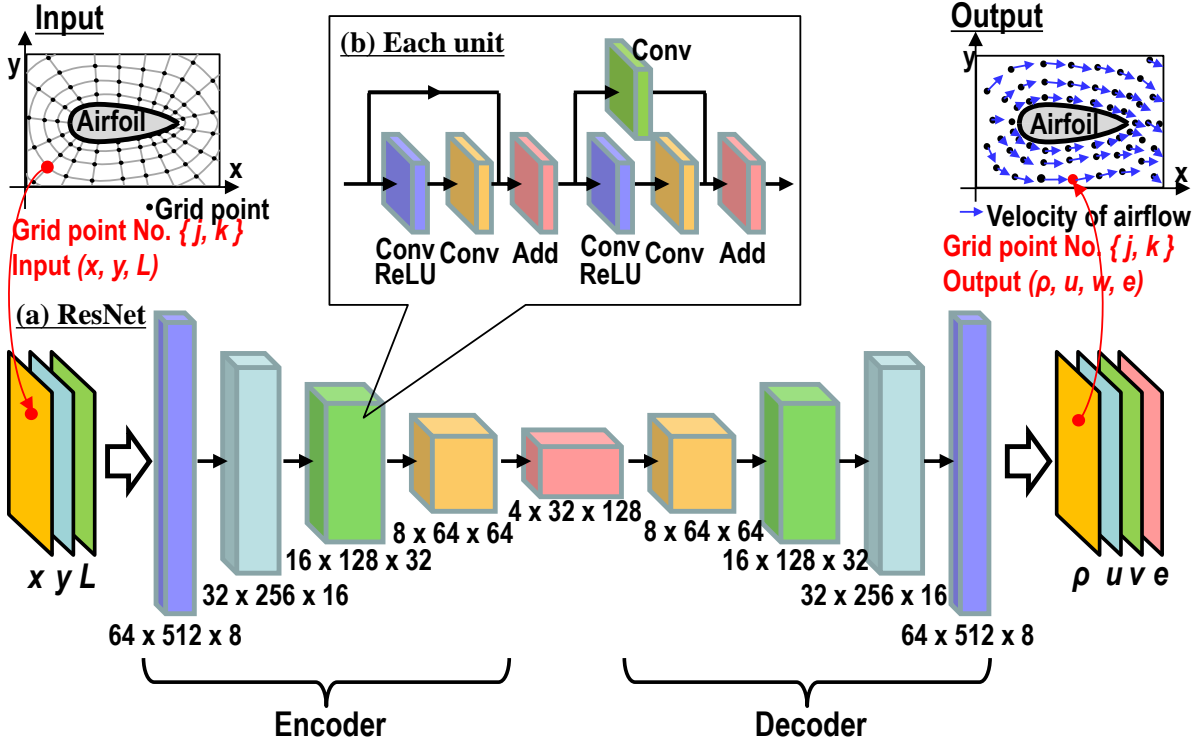

Fig. 3. Architecture of the DNN used to infer the flow field. The airfoil shape and the placement of the grid points are used to input of the DNN. The DNN is composed of ResNet. The output is flow variables such as density "ρ", velocity "u", "v", and energy "e" on each Grid point.

The grid points concerning the shape were treated as pixels in the image. The grid point $\{j, k\}$ is allocated to pixel $\{j, k\}$ in the image. The input information of the grid point $\{j, k\}$ is the $x$ coordinate, $y$ coordinate, and distance from the airfoil $L$. The output information of the grid point is the value of the flow variables, such as density "$\rho$"; velocity of the $x, y$ directions "$u$", "$v$"; and energy "$e$".

### B. Network Design and Architecture

A residual neural network (ResNet) was employed as the DNN architecture to infer the flow field. ResNet has shown remarkable performance in image classification [29, 30]. The architecture of the DNN utilized to infer the flow field is shown in Fig. 3. The DNN consists of an encoder and a decoder. In the encoder part, the features of the input shape were extracted. In the decoder part, the flow field was inferred based on the extracted features. The convolutional layers are used to extract the features of the input image. The encoder and decoder are composed of four units. Each unit includes convolutional and shortcut connections as shown in Fig.3(b).

### C. Control ID Number vs Paper Number

The loss function used to train the DNN was customized to infer the flow field for the nonuniform allocation of grid points. To train DNN, the MSE is often applied as a loss function in regression problem [36]. Actually, the MSE is used as a loss function in previous research [26]. The equation for the loss function using the MSE is expressed as follows:

$$(Loss\ 1) = mean\left(\sum_{j}^{jmax}\sum_{k}^{kmax}\left(A_p(\rho, u, v, e) - A_t(\rho, u, v, e)\right)^2\right), \quad (1)$$

Where "$\{j, k\}$" is a grid number and "$jmax$", "$kmax$" is the number of grid points. $A_p(\rho, u, v, e)$ means the inferred values of the physical quantities. $A_t(\rho, u, v, e)$ means the ground truth values of the physical quantities. $At(\rho, u, v, e)$

calculated using CFD. "$\rho$", "$u$", "$v$", and "$e$" are the physical quantities that represent the flow states. All errors at every grid point are suppressed equally with this loss function.

However, CFD usually uses nonuniform allocation of grid points. Therefore, the associated area, depending on each grid point, is also nonuniform. The physical quantities of grid points in a sparse area affect large areas. As a result, the error on these grid points is proportionally large. This error will influence entire flow field including in area of around airfoil. To reduce this effect, the error value is multiplied by the size of the associated area related to each grid point, which is realized through the customization of the loss function used in the DNN training. The customized loss function is expressed as follows:

$$(Loss\ 2) = mean\left(\sum_{j}^{jmax}\sum_{k}^{kmax} S(j,k)\left(A_p(\rho,u,v,e) - A_t(\rho,u,v,e)\right)^2\right), \quad (2)$$

where $S(j, k)$ is the size of the associated area for each grid point. The effect of the nonuniform allocation of the grid point is canceled by multiplying the size of the associated area by the square error of each grid point. As a result, the accuracy of the DNN inference will be expected to improve further from the perspective of the entire flow field. Here, Jacobian $J$, used to convert to generalized coordinate system ($j, k$) from Cartesian coordinates ($x, y$), has a relationship with this size of the associated area. The Jacobian $J$ is expressed as follows:

$$J = 1/\left(\frac{\partial x}{\partial j}\frac{\partial y}{\partial k} - \frac{\partial x}{\partial l}\frac{\partial y}{\partial k}\right). \quad (3)$$

Here, the "$1/J$" is generally equivalent to the size of the associated area of the unit grid point. Therefore, "$1/J$" is applied as $S(j, k)$ in this study. Therefore, the error on the grid point located in the sparse region also has a significant influence proportional to the size of associated region. This influence is not considered in the previous research [26]. Training DNN by considering this influence, the accuracy of the DNN inference will be expected to improve further from the perspective of the entire flow field.

## IV. Experimental Setup

The objective of this study is to suppress the computational cost of CFD utilized for design optimization. Design optimization problem of two-dimensional airfoil shape using PARSEC parameters [17, 18] was used as the sample problem this time. Therefore, the airfoil shapes of the design candidates were produced based on the PARSEC parameter in this case. The airfoil flow computations are performed at a Reynolds number of 3,000,000 and Mach 0.30 in this study. The evaluation was performed at an AOA of 2 degrees.

The simulator calculates the flow field by solving the compressible Navier-Stokes equations. In this study, LANS3D was employed as the CFD simulator [31]. We employed upwind-biased 2nd-order differencing [32], the lower–upper symmetric Gauss–Seidel scheme [33], and the turbulence model of Baldwin and Lomax [34].

The DNN was implemented using Keras [35] and a TensorFlow GPU 1.13 backend. The CFD simulation was run on an Intel Xeon E5-2690 (2.90 GHz). DNN training was performed on the dual NVIDIA Tesla P100 GPU. The DNN infers flow field of steady solution from the information of CFD grid. Therefore, the DNN is trained using CFD grids of airfoil shape and the flow fields of steady solution. The dataset consisting of the CFD grid and the flow fields is created using traditional CFD simulator. The sample problem is design optimization problem using PARSEC parameters. Therefore, we prepared the design candidate using a random number for the PARSEC parameter. A total of 210 designs were prepared as the airfoil shapes of the design candidates. A total of 180 designs of 210 designs were used to train the DNN. The training dataset comprised 180 samples, including 36 samples for validation. The batch sizes for the training were 8, and the learning rate was 1.0e-3. In the training stage, the MSEs or customized MSE are employed as the loss functions to compare the results.

The other 30 designs were applied to evaluate both the DNN part and the performance of the CFD with the DNN. The result of the traditional CFD is the baseline. Therefore, these 30 designs were evaluated using traditional CFD. The accuracy of the DNN inference was evaluated using the result of the traditional CFD. The reduction in the computational cost of our CFD with the DNN was also evaluated using the results of the traditional CFD.

## V. Computational Time of CFD for One Design

The performance of our CFD incorporated with DNN inference was evaluated. First, we evaluated the characteristics of the DNN unit. We then evaluated the entire CFD model incorporated with the DNN inference. 30 design candidates were evaluated to confirm the effects of individual differences. First, the evaluation results of one design candidate are shown as a sample. Second, the evaluation results of the 30 design candidates are shown.

### A. Characteristics of DNN Unit

We evaluated the flow field of the DNN inference against that of the steady solution. The computational cost of the main CFD depends on the difference between the flow field of the DNN inference and that of the steady solution. The accuracy of DNN inference should be improved to reduce the computational cost. The objective of this study is to suppress the computational cost of CFD by using the flow field inferred by the DNN. Therefore, the accuracy of DNN inference is an important criterion.

#### 1. DNN Characteristics Depending on the Input Information

We evaluated the characteristics of the DNN by inferring the flow field from the shape of the design candidate. The DNN characteristics, depending on the input information, were evaluated to determine the input variables. The DNN is trained using MSE, as in the reported research [26]. The value of the MSE is also used as a criterion for accuracy. We evaluated the effect of using both "$x$", "$y$" coordinates and distance "$L$" as the input information. Figure 4 shows the inferred flow field, depending on the input information. The flow field is inferred only from the distance from the surface of airfoil "$L$" (Fig. 4(a)), only from the positions of coordinates of the grid point "$x$", "$y$" (Fig. 4(b)), and from both the position of the coordinates and the distance (Fig. 4(c)). The MSE value is 6.2e-3, with the flow field inferred only from the distance. The MSE value is 3.5e-3, with the flow field inferred only from the coordinates. Here, the MSE value is reduced to 2.3e-3 with the flow field inferred from both the coordinates and distance. We confirm the effect of using both the coordinates and distance as inputs to the DNN.

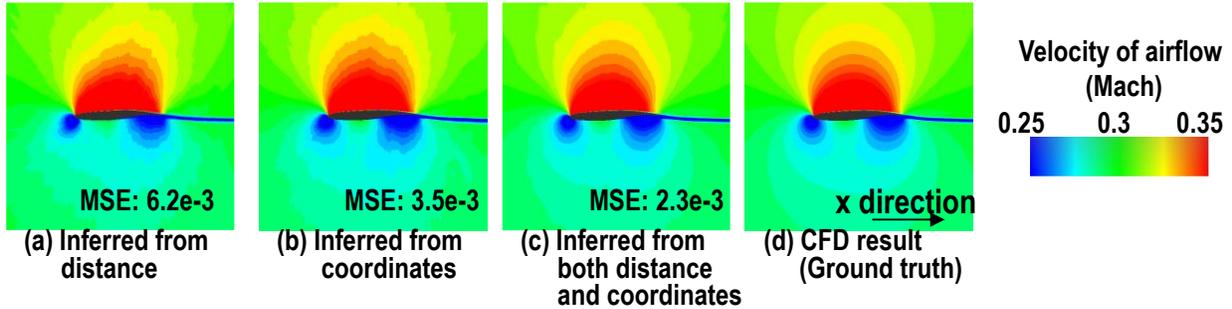

**Fig. 4. Inferred flow field depending on the input information. The figure shows the flow velocity of x-direction. The flow field is inferred from the distance from the airfoil (a), from the position on coordinates (b), both the position on coordinates and the distance (c), and the CFD result (d).**

#### 2. Characteristics of DNN Trained by the MSE (DNN 1)

We applied both the coordinates and distance as inputs of the DNN. The MSE (equation (1)) is utilized as a loss function to train the DNN, first (DNN 1). Figure 5 (a), (d) shows the flow field inferred by the DNN 1. Here, the ground truth is the result of traditional CFD. Figure 5 (c) and (f) show the flow field of traditional CFD. The flow field inferred by DNN 1 is close to that of CFD in the region closest to the airfoil. However, in the region far from the airfoil, some differences were observed between the flow field inferred by DNN 1 and flow field calculated by CFD.

The MSE is employed as a loss function to train the DNN. In this case, the DNN was trained to equally suppress errors in physical quantities on each grid point. Therefore, the inference errors on each grid were almost equal. However, the size of the associated area for each grid point was not uniform. When the size of the associated area is large, the inference error is equivalently large. The grid points are sparsely allocated in the region far from the airfoil, that is, the associated area of this grid point is large. Thus, the inference error far from the airfoil is substantially large, which explains why the inference error far from the airfoil is large. The effect of the size of the associated area for each grid point needs to be compensated to overcome this issue.

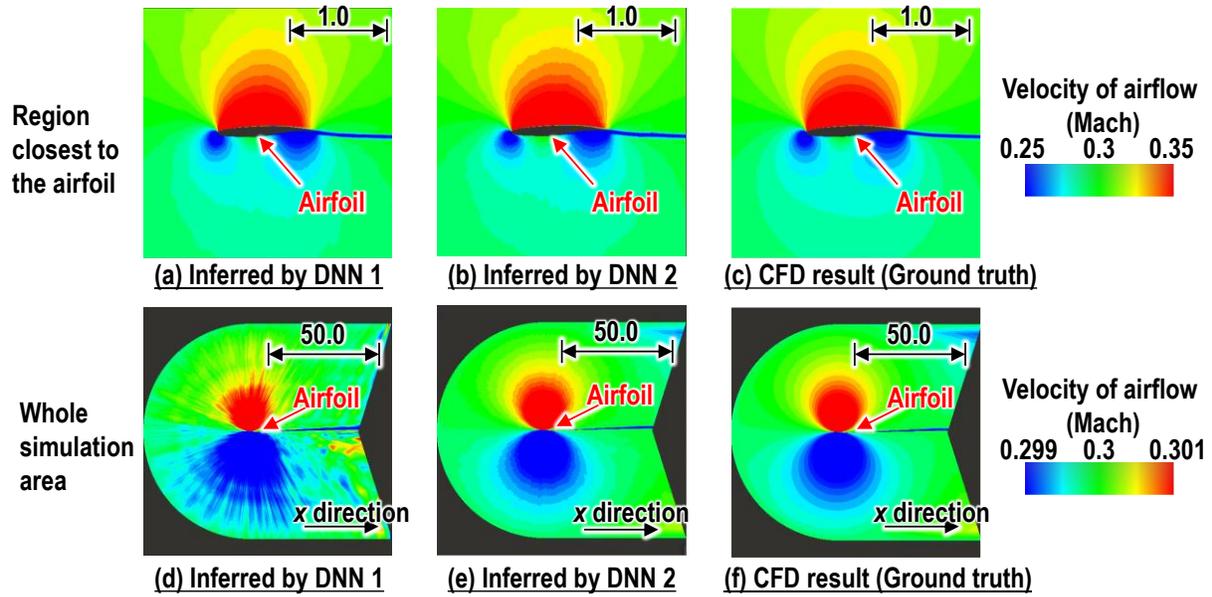

**Fig. 5.** Inferred flow field with DNN and the CFD result. The figure shows the flow velocity of x-direction. Scale of the distance is nondimensionalized by the chord length of airfoil. The flow fields closest to the surface of the airfoil (a), (b), (c) and the whole area (d), (e), (f) are shown.

**3. Characteristics of DNN Trained by the Customized MSE (DNN 2)**

We applied a customized loss function to improve the flow field in the region far from the airfoil. The customized MSE (equation (2)) instead of the MSE is applied to train the DNN (DNN 2). Figure 5 (b) and (e) show the flow field inferred by DNN 2. The issue caused by the size of the associated area for each grid point was compensated by this method. The flow field inferred by DNN 2 is close to that of the CFD in the region closest to the airfoil and far from the airfoil.

**(a) Traditional CFD**

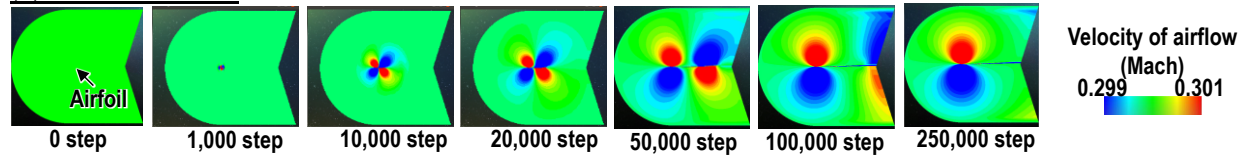

**(b) CFD with our DNN1**

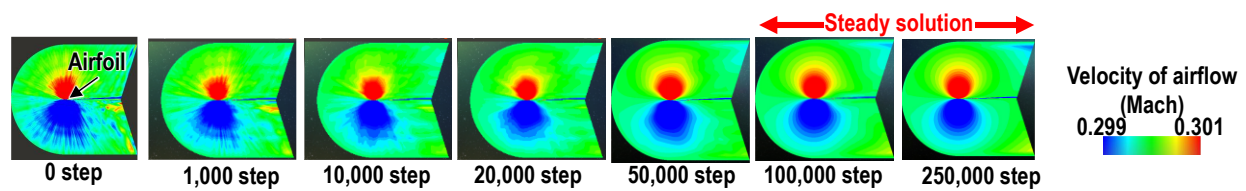

**(c) CFD with our DNN2**

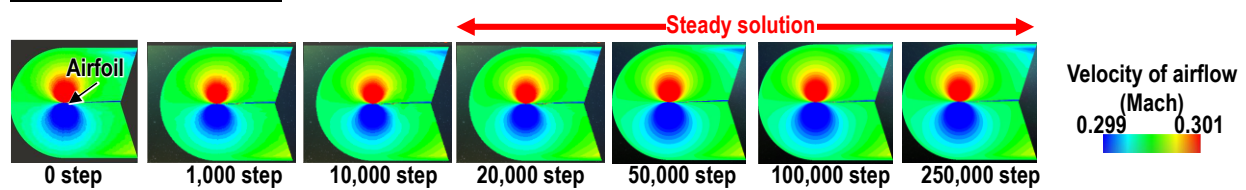

**Fig. 6.** Flow field depending on the iterative number of the CFD (a) Traditional CFD, (b) CFD with our DNN1, and (c) CFD with our DNN2. The figure shows the flow velocity of x-direction. The flow field of 250,000 step indicates the steady state flow field.

## B. Evaluation Results of Our CFD

The feature of our CFD simulation enables the suppression of the computational cost. And yet, there is no difference between the results of the traditional CFD and those of our CFD. Therefore, we obtained the results using traditional CFD and our CFD. We then confirm that the results are the same. Subsequently, we evaluated the effect of suppressing the computational cost of our CFD.

### 1. Accuracy of the Results of Our CFD Simulation

We evaluated the results using the traditional CFD, our CFD with DNN 1, and our CFD with DNN 2. We compared the flow field depending on the iterative number of the main CFD. The flow field of one design candidate is shown in Fig. 6. The flow field at 250,000 steps with traditional CFD is the actual value of the steady solution. As shown in the flow field at 250,000 steps, the flow field of the steady solution is the same regardless of the CFD. On the other hand, the iterative number required to achieve a steady solution depends on the CFD.

We confirmed that the results were the same regardless of the CFD; this was verified using the value of the objective function. The value of $C_D$ at steady solution is one of the major objective function of the design candidate [23]. Therefore, the difference against the value of $C_D$ at the steady solution was employed as a criterion in this study. Figure 7 shows the $C_D$ values calculated from each flow field against the number of iterative calculations. The value of $C_D$ at the steady solution solved by traditional CFD is treated as the actual value. As shown in Fig. 7, the value of $C_D$ also converges to the same value regardless of the CFD simulation.

### 2. Suppression of the Computational Cost of Our CFD Simulation

We evaluated the effect of reducing the computational cost of our CFD simulation. The difference against the value of $C_D$ at steady solution was applied as a criterion. The value of $C_D$ at 250,000 steps solved by the traditional CFD simulation was used as a reference for an actual value. The value of $C_D$, depending on the number of iterations, is calculated as shown in Fig. 7. The value of $C_D$ below $\pm 0.00005$ (1 count) against the reference was regarded as a steady solution. The computational cost is proportional to the number of iterations required to achieve this condition. Therefore, the number of iterations required to achieve this condition is defined as the computational cost in this section. In case the sample design candidate is evaluated in Fig. 7, the number of iterations is 121,000 steps with the traditional CFD, 80,400 steps with the CFD with DNN 1, and 9,700 steps with the CFD with DNN 2.

The number of iterations depends on the design candidate. Therefore, the number of iterations was evaluated for all 30 design candidates. Figure 8 shows the results of the six examples. As shown in these results, our CFD with the DNN is effective in reducing the computational cost for various design candidates. The number of iterations required to realize convergence for the 30 design candidates is evaluated. The average number of iterations was 112,600 for traditional CFD, 65,600 for CFD with DNN 1, and 29,000 for CFD with DNN 2. The CFD with DNN 2 is accelerated by a factor of 3.9× against the traditional CFD. The CFD with DNN 2 is accelerated by a factor of 2.3× against the CFD with the DNN 1 inference.

## C. Comparison with the Previous Research

Our CFD with the DNN converges to the same results as the traditional CFD. This finding also indicates that our CFD with the DNN converges to the same value as the previous CFD introduced in a previous study [26]. Therefore, this result also indicates that we succeeded in eliminating the warm-up CFD stage using our CFD.

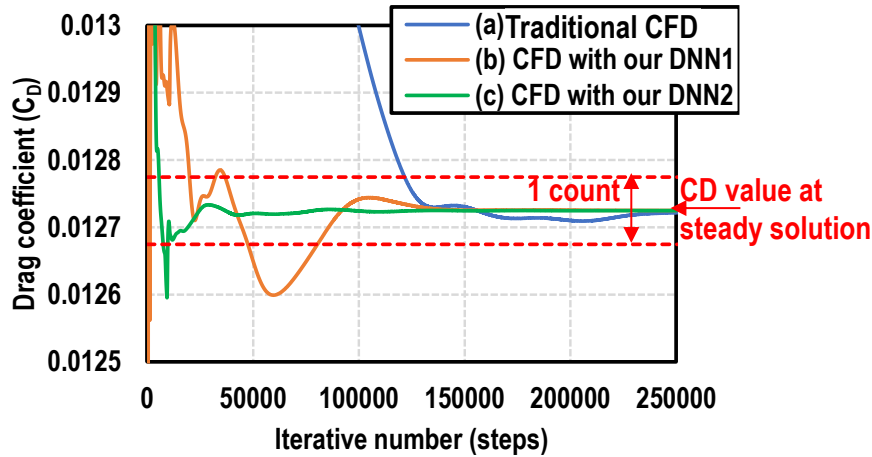

**Fig. 7. CD value calculated from the Flow field depending on the iterative number of the CFD (a) Traditional CFD, (b) CFD with our DNN1, and (c) CFD with our DNN2.**

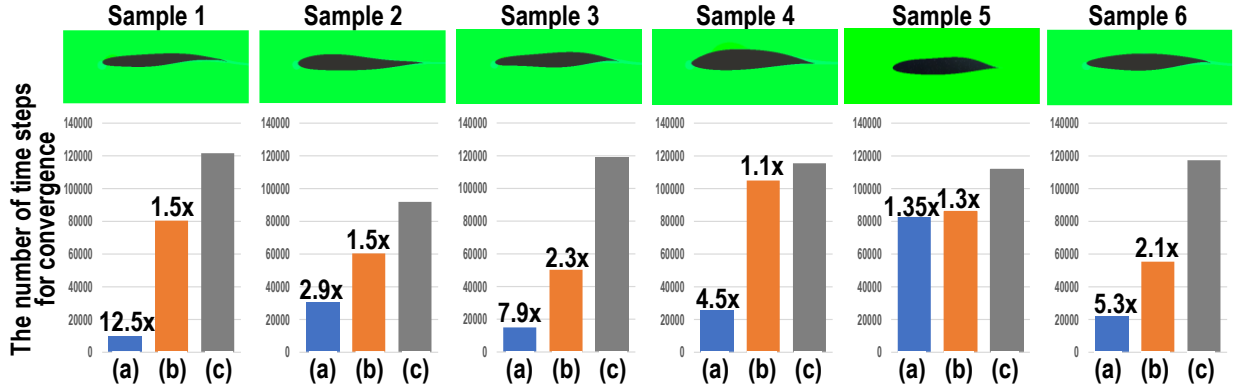

**Fig. 8.** The number of the time steps to the convergence of various airfoil shape. The upper figures show the shapes of evaluation performed, and the lower figures show the evaluation results of each shape. (a) CFD with our DNN 2, (b) CFD with our DNN 1, (c) CFD without DNN (Traditional CFD)

## VI. Estimate Total Computational Time for Design Optimization

We estimated the total computational time for the case in which design optimization was performed using our CFD with the DNN. To apply our CFD with the DNN, the DNN training process needs to be performed during the entire process. Therefore, the overhead time of this process must be considered. The estimation is performed with the case of design optimization with 1,000 design candidates as an example. DNN training was performed with 180 design candidates in this study. Therefore, the entire flow of design optimization is listed as follows:

(1) 180 design candidates were evaluated using traditional CFD. These 180 results were used for both the evaluation of the objective function and DNN training data.
(2) DNN training was performed using these 180 results.
(3) The remaining 820 design candidates were evaluated using our CFD with the DNN.
(4) A suitable design was selected based on the results of all 1,000 design candidates.

### A. Execution Time of Each Process

The execution time of our proposed method is evaluated. The execution time per design candidate is 10.5 hours at 1 core for traditional CFDs. The execution time per design candidate is 2.7 hour at 1 core for our CFD with DNN 2. It takes 9.5 hour to train the DNN. It takes less than 1 minute to predict the flow field by the trained DNN with a 1 core central processing unit (CPU). The execution time of DNN inference is negligible compared to that of CFD. Therefore, the simulation times of the CFD part are regarded as the total execution time of our CFD with DNN 2.

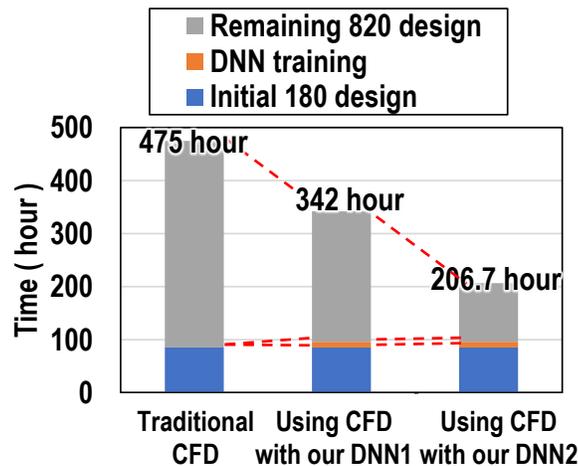

**Fig. 9.** Comparison result of estimate total computational time

### B. Estimate Time of Entire Workflow of Optimization

We estimated the total computational time for the case of the 20 core CPU. The estimated times are shown in Fig. 9. With the 20 core CPU, it takes 475 hours with traditional CFD. And, it takes 206.7 hours with our CFD for the entire workflow.

## VII. Conclusions and Future Work

For expensive real-world aerodynamic design optimizations, a new approach for acceleration of flow simulations using DNN is proposed. In this approach, plausible initial flow condition is predicted by DNN to significantly reduce number of time steps required for convergence to the steady state. DNN is trained with flow simulations prior to aerodynamic design optimization where coordinates and distance from the wall surface of all grid points for flow simulation are the input and physical properties such as static pressure and flow velocity are the output.

In this study, proposed approach was evaluated for an aerodynamic airfoil shape design optimization problem. Here, 174 flow simulation results of randomly generated airfoil shapes were prepared. DNN was trained with 144 flow simulation results and the speed up of the proposed approach was evaluated with 30 flow simulation results. The results showed that 3.9x speed up is achieved in average compared with standard flow simulation without initial flow prediction by DNN.

Though the proposed approach requires additional flow simulations before aerodynamic design optimization for training DNN, in total, significant speed up can be achieved for aerodynamic design optimization. For example, if the proposed approach is applied to an aerodynamic design optimization that requires 10,000 flow simulations and 180 additional flow simulations for training of DNN, the speed up is estimated as 10,000/(10,000/3.9+180)=3.64.